\documentclass[aps,twocolumn,floatfix,prl]{revtex4-1}
\usepackage{latexsym,amssymb,graphicx,amsmath,epsfig,calc,times}

\usepackage[usenames]{color}

\citeindextrue

\def\btt1{{\tt$\backslash$\string1}}%

\def\AmS{{\protect\the\textfont2
        A\kern-.1667em\lower.5ex\hbox{M}\kern-.125emS}}

\def\rhotwod{\rho_\text{2d}}

\def\Nr{N_\text{r}}
\def\ps{p_\text{s}}
\def\posm{p_\text{s}}

\definecolor{Blue}{rgb}{0,0.0,1.0}
\definecolor{Red}{rgb}{1.0,0.0,0.0}
\definecolor{Green}{rgb}{0.0,0.35,0.0}
\definecolor{Grey}{rgb}{0.5,0.5,0.5}
\newcommand{\vex}{v_\text{ex}}
\newcommand{\rhom}{\rho_\text{m}}
\newcommand{\kt}{k_\text{B}T}

\newcommand{\Rg}{R_\text{g}}

\newcommand{\fp}{f_\text{pr}}

\newcommand{\ds}{d_\text{s}}

\citeindextrue

\begin{document}
\title{Self-assembly of 2D membranes from mixtures of hard rods and depleting polymers}
\author{Yasheng Yang}
\author{Edward Barry}
\author{Zvonimir Dogic}
\author{Michael F. Hagan}
\affiliation{Department of Physics, Brandeis University, Waltham, MA, 02454}
\email{hagan@brandeis.edu}
\date{\today}

\begin{abstract}
We elucidate the molecular forces leading to assembly of two dimensional membrane-like structures composed of a one rod-length thick monolayer of aligned rods from an immiscible suspension of hard rods and depleting polymers. We perform simulations which predict that monolayer membranes are thermodynamically stable above a critical rod aspect ratio and below a critical depletion interaction length scale. Outside of these conditions alternative structures such as stacked smectic columns or nematic droplets are thermodynamically stable. These predictions are confirmed using an experimental model system of virus rod-like molecules and non-adsorbing polymer. Our work demonstrates that collective molecular protrusion fluctuations alone are sufficient to stabilize membranes composed of homogenous rods with simple excluded volume interactions.
\end{abstract}

\maketitle

Colloidal membranes are two dimensional (2D) surfaces composed of a one rod-length thick monolayer of aligned nanorods. Equilibrium formation of such structures requires assembly to readily propagate in two dimensions while self-limiting the third. Previous approaches towards assembly of colloidal membranes utilized chemically heterogeneous rods that mimic the dichotomic structures of amphiphilic molecules comprising biological membranes~\cite{Park2004}. Here, we use computer simulations to demonstrate that structurally and chemically homogeneous hard rods can form equilibrium monolayers in the presence of depletant molecules, suggesting that geometry as well as chemical heterogeneity can be used to design assembly pathways of self-limited structures. Furthermore, we discover bounds on the molecular parameters that support formation of equilibrium membranes. These results have fundamental as well as practical significance. Extensive research has shown that hard particle fluids undergo entropy-driven assembly into a myriad of 3D structures~\cite{Bolhuis1997,Onsager1949,Chandler1983,Pusey1986}. Our work demonstrates that entropic forces can also drive formation of 2D structures. From a practical perspective, equilibrium colloidal membranes may enable manufacture of inexpensive and easily scalable optoelectronic devices~\cite{Baker2010}.

\begin{figure}
\epsfxsize=0.79\columnwidth\epsfbox{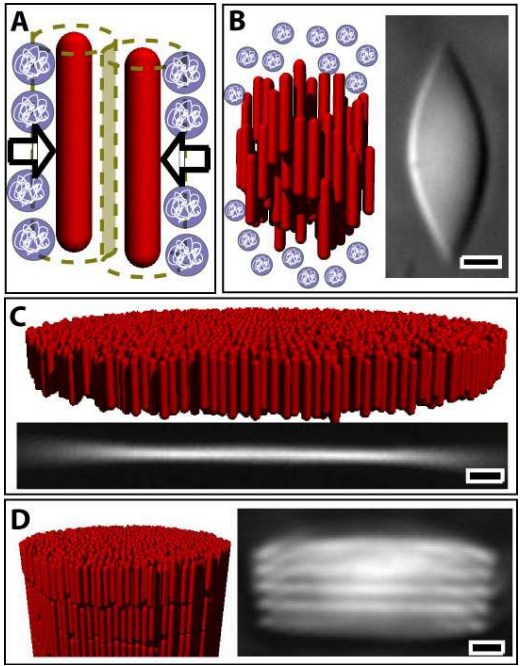}
\caption{Schematic illustrations and optical micrographs of the self-assembled structures observed in suspensions of the filamentous virus {\it fd} and non-adsorbing polymer~\cite{Barry2010}. {\bf A)} Non-adsorbing polymer induces effective attractive interactions between rods. {\bf B)} DIC micrograph and schematic of a nematic tactoid formed at low depletant concentration. {\bf C)} At intermediate depletant concentrations, rod-like viruses condense into macroscopic one rod-length 2D fluid-like membranes. {\bf D)} At high depeltant concentration, membranes stack on top of one another, forming smectic filaments. All scale bars are 5$\mu$m.
\label{fig:experiment} }
\end{figure}

Our study is motivated by recent experiments on suspensions of monodisperse rod-like colloidal viruses and the non-adsorbing polymer Dextran~\cite{Barry2010} (Fig.~\ref{fig:experiment}). \emph{fd} viruses alone approximate the behavior of homogenous rods interacting with repulsive hard-core interactions~\cite{Purdy2003}. The polymer induces an entropy-driven attractive (depletion) potential between the rods, the strength and range of which can be tuned by changing the polymer concentration and radius of gyration respectively (Fig.\ref{fig:experiment}A)~\cite{Asakura1954}. At high polymer concentrations viruses condense into smectic-like stacks of 2D membranes (Fig.\ref{fig:experiment}D)~\cite{Frenkel2002}. With decreasing polymer concentration (attraction strength), individual 2D monolayers (membranes) within a smectic filament unbind, indicating that the membrane-membrane interaction switches from attractive to repulsive~\cite{Barry2010} (Fig.\ref{fig:experiment}C).

Understanding the molecular origin of the repulsive membrane-membrane interactions is our primary goal. Experiments revealed significant protrusions of rods from isolated colloidal membranes, the magnitude of which could be tuned by changing the concentration of non-adsorbing polymer~\cite{Barry2010}. In contrast, these fluctuations were suppressed in stacked membranes. It was proposed that the entropy penalty associated with suppressing protrusion fluctuations of individual rods leads to repulsive interactions that stabilize isolated membranes under moderate osmotic pressure~\cite{Israelachvili1992}. However, other plausible factors could also stabilize membranes, including attractive interactions between virus tips and depletant, repulsions due to bending (Helfrich) modes, or kinetic trapping of membrane intermediates. To elucidate these issues, we develop a computational model which demonstrates that protrusion interactions alone are sufficient to stabilize membranes in equilibrium. In contrast to the previous model which considered only protrusions of isolated rods~\cite{Barry2010}, our work indicates that collective protrusion undulations dominate repulsive interactions between membranes. Surprisingly, the simulations predict that membranes are stable only for a certain range of rod aspect ratios and depletant sizes. We experimentally confirm the latter prediction.

We model the {\it fd} rods as hard spherocylinders with diameter $\sigma$ and length $L$. The non-adsorbing polymer is modeled as ghost spheres of diameter $\delta$ which freely interpenetrate one another but behave as hard spheres when interacting with rods~\cite{Asakura1958}. Compared with an effective pair potential approach, this model accounts for multi-rod interactions induced by polymers~\cite{Savenko2006,Patti2009,Cuetos2010}.
We perform Metropolis Monte Carlo (MC) with periodic boundary conditions~\cite{Frenkel2002a}. The total number of rods $\Nr$ is fixed, the sphere osmotic pressure $\ps$ is set by insertion/deletion moves, and constant pressure is maintained in the $xy$ plane by performing volume-change moves, while the box size is fixed in $z$ direction \footnote{ The number of rods is $\Nr=512$, except for simulations that examine finite size effects, in which $128\le \Nr\le 1152$, and with orientational fluctuations, in which $\Nr=1560$.  For free energy calculations with orientational fluctuations and large aspect ratios, rods are allowed to interact with multiple periodic images of other rods (following Ref. \cite{Bolhuis1997}) and orientational fluctuations beyond a maximum angle are rejected to prevent any rod from interacting with itself.  The maximum allowed angle is well beyond typical orientational fluctuations since rods in membranes are nearly aligned. Varying the maximal allowed angle showed that the constraint did not affect the free energy.}. Simulation results are reported with $\sigma$ as the unit of length, $k_BT$ as the unit of energy, and $k_BT\sigma^{-3}$ as the unit of pressure.

Membrane-like structures have two generic repulsive interactions of distinct origin which dominate at different separation lengthscales. At large separations, slowly decaying low energy bending (Helfrich) modes dominate~\cite{Helfrich1984}. In contrast, at separations comparable to the monolayer thickness (the rod length), protrusions of molecules from the membrane surface generate the primary repulsive force~\cite{Goetz1999}. We expect that long ranged bending modes provide a negligible contribution to the stabilization of colloidal membranes for the following reasons. First, the range of the depletion attraction that balances repulsive forces to drive membrane stacking is comparable to the depletant size. On this scale protrusion repulsions dominate. Second, the bending modes involve deviations of rods from the preferred direction and their interaction strength scales with bending modulus as $f_\text{bend} \sim \kappa_\text{c}^{-1}$ \cite{Helfrich1978,Safran1994}. The large bending modulus measured for {\it fd} membranes $\kappa_\text{c}=150 \kt$ thus results in very weak Helfrich repulsions~\cite{Barry2010,Safran1994}. Consistent with these arguments, a theoretical calculation (SI Fig. 5) shows that the protusion interactions in simulated membranes exceed the strength of the Helfrich interactions by four orders of magnitude at relevant separations.

\begin{figure}
\epsfxsize=0.95\columnwidth\epsfbox{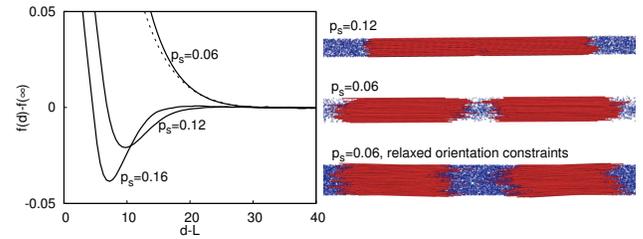}
\caption{(left) Free energy per rod of an interacting membrane pair, $f(d)-f(\infty)$, plotted as a function of membrane surface separation $d-L$, shown for three different depletant concentrations indicated by values of $\ps$, with sphere diameter $\delta=1.5$ and aspect ratio $L=100$.   The dashed line is the free energy calculated with orientational fluctuations at $\ps=0.06$. (right) Snapshots of two membranes from unbiased trajectories. (top) Membranes attract at $\ps=0.12$. (middle, bottom) Snapshots for $\ps=0.06$ from simulations with (middle) parallel rods and (bottom) rods with orientation fluctuations. In both cases membranes drift apart, indicative of a repulsive potential.
\label{fig:wham}
}
\end{figure}

Based on the preeminence of protrusion modes, which do not involve rod tilting, in most simulations we restrict spherocylinder orientations to be perfectly aligned along the $z$ direction. This simplification greatly enhances computational efficiency, allowing us to extensively map the phase diagram as a function of all relevant molecular parameters. Our approximation is justified by Fig. 5 in the SI and the fact that simulations in which the fixed orientation constraint is relaxed predict similar phase behavior and membrane-membrane interactions (e.g. Fig.~\ref{fig:wham}).

\emph{Membrane-membrane interaction potential.}
We first use umbrella sampling~\cite{Frenkel2002a} to measure the free energy per rod $f$, as a function of the separation between the centers of mass of two membranes, $d$ (Fig.~\ref{fig:wham}). At low osmotic pressures (e.g. $\ps=0.06$), $f(d)-f(\infty)$ has no attractive region sufficient to overcome translational entropy; i.e., the stacking of disks is suppressed and the isolated colloidal membrane phase is stable.
For larger osmotic pressures ($\ps\gtrsim0.08$), the free energy has a substantial minimum at finite membrane separations, signifying that membranes will stack to form the smectic-like columns. Consistent with these free energy results, unbiased simulations for these parameters resulted in two membranes which were respectively isolated and stacked at low and high osmotic pressures, as shown in Fig.~\ref{fig:wham} (right).  The free energy and a  representative snapshot are also shown for rods with orientational fluctuations at $\ps=0.06$. Note that isolated membranes are stable and the interaction free energy is comparable to the case with parallel rods;  the repulsion is slightly weaker with orientational fluctuations because they decrease the equilibrium areal rod density.

\begin{figure*}
\epsfxsize=\textwidth\epsfbox{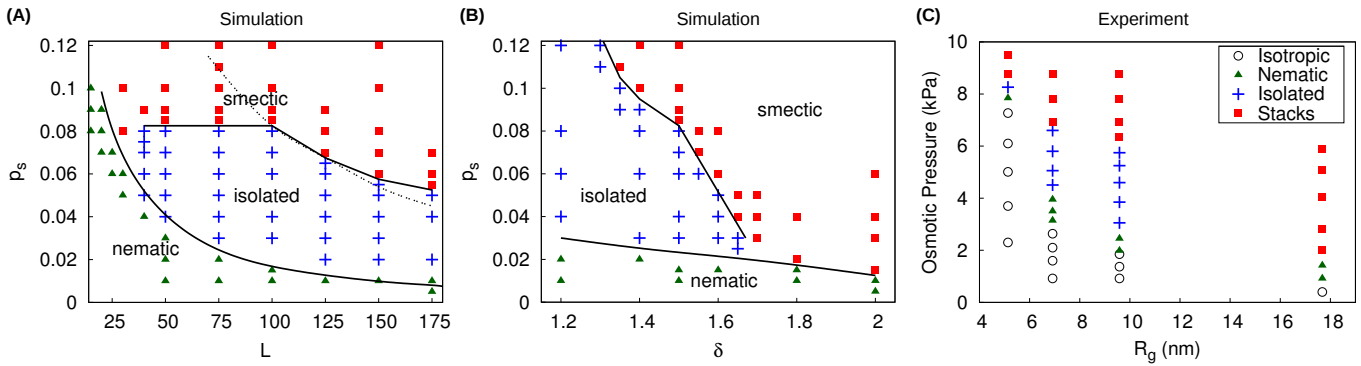}
\caption{Phase diagrams from simulation and experiment. Triangles $\blacktriangle$ denote denote parameters that lead to nematic configurations, + symbols correspond to isolated membranes, and $\blacksquare$ symbols correspond to smectic layers. {\bf (A),(B)} Phase diagrams determined from simulations for varying osmotic pressure $\ps$ and {\bf (A)} aspect ratio $L$  with sphere diameter (polymer radius of gyration) $\delta=1.5$, and {\bf (B)} varying sphere diameter with $L=100$. The solid lines identify the  the isolated membrane/smectic and nematic/isolated membrane phase boundaries. They are fit by eye to simulation results except for the  nematic/isolated membrane boundary in {\bf(A)}, which is a theoretical prediction \cite{Yang2011}. {\bf(C)} The experimental phase diagram corresponding to {\bf (B)} using mixtures of \textit{fd} viruses and PEG/PEO polymers. The final concentration of viruses was fixed at 5mg/mL and both polymer concentration and molecular weight were varied to change osmotic pressure and polymer radius of gyration, $R_g$, respectively. As noted in the text, chiral structures such as helical ribbons which appear near the nematic and isolated membrane boundary are not shown.
 \label{fig:phasesSphereDiameter}
}
\end{figure*}

\emph{Phase diagram.} We computed the equilibrium phase behavior as a function of osmotic pressure, rod aspect ratio, and sphere diameter as follows  (Fig.~\ref{fig:phasesSphereDiameter}). To identify the nematic-membrane phase boundary, we performed separate unbiased simulations starting from initial conditions in which (1) rods have random positions and (2) rods are aligned in a flat layer. For all results shown, the simulation outcomes were independent of initial conditions. To identify the transition from membranes to smectic filaments, a parameter set was considered to yield smectic layers if the total free energy of the attractive basin in the membrane-membrane interaction potential satisfies $F \le F_0=\kt \ln\rho_\text{m} v_0$ with $\exp(-\beta F)=\int_{f(s)<0}ds\exp(-2\beta Mf(s))$, $M$ the number of rods in one membrane, $v_0$ a standard state volume, and $\rhom$ a membrane concentration. A finite value of $F_0$ accounts for membrane translational entropy. We roughly estimate $M=10^4$ and $\rhom v_0 = 10^{-8}$ from the experimental conditions; the location of the phase boundary is not sensitive to the value of $\rhom v_0$.

Fig.~\ref{fig:phasesSphereDiameter}A illustrates the location of the equilibrium nematic phase, isolated membranes, and smectic stacks as a function of rod aspect ratio and depletant concentration. Interestingly, isolated membranes are thermodynamically stable over a significant span of osmotic pressures, but only for rods with aspect ratios larger than $L=30$. Simulations with orientational fluctuations also indicate a minimum aspect ratio for stable membranes, which is somewhat larger. These predictions are consistent with previous simulations of rods with $L=5$ that did not find equilibrium monolayers \cite{Patti2009}. The disappearance of the isolated membrane phase for shorter rods arises from the interplay between the geometry of rod-like particles and attractive depletion interactions. Since the strength of the attractive interaction between two rods scales linearly with rod length, increasing the rod length lowers the osmotic pressure associated with the nematic to membrane transition. On the other hand, the transition from isolated membranes to smectic filaments is determined by the roughness of colloidal membranes, which is independent of rod length but decreases with increasing depletant concentration.  Based on this argument,  the location of the transition between colloidal membranes and smectic filaments should be independent of rod length, which is indeed observed for rod lengths between 30 and 100. For longer rods the location of the transition slightly decreases with increasing rod length, due to 2D crystallization of rods within membranes (see the SI for details of membrane crystallization and a determination that finite size effects do not affect the results). At a critical rod length the nematic-membrane phase boundary intersects the membrane-smectic filament phase boundary, ending the equilibrium membrane phase.

Fig. \ref{fig:phasesSphereDiameter}B reveals that the depletant size $\delta$ also influences the topology of the phase diagram. For $\delta > 1.7$ colloidal membranes are unstable at all osmotic pressures and there is a direct transition from the nematic phase to smectic filaments. In contrast, for $\delta < 1.7$ colloidal membranes are the equilibrium phase at intermediate depletant concentrations between a low osmotic pressure nematic phase and high osmotic pressure smectic filament phase. Decreasing the depletant size further below this critical value significantly expands the range of osmotic pressures for which colloidal membranes are stable. These results can be understood as follows. Increasing the depletant size expands the effective range of the attractive potential between two colloidal membranes, which in turn requires longer range repulsive interactions to stabilize colloidal membranes. For large enough depletant molecules, the repulsive protrusion interactions are not sufficiently long-ranged to overcome the attractive potential and colloidal membranes become unstable for all osmotic pressures.

\begin{figure}
\epsfxsize=0.49\columnwidth\epsfbox{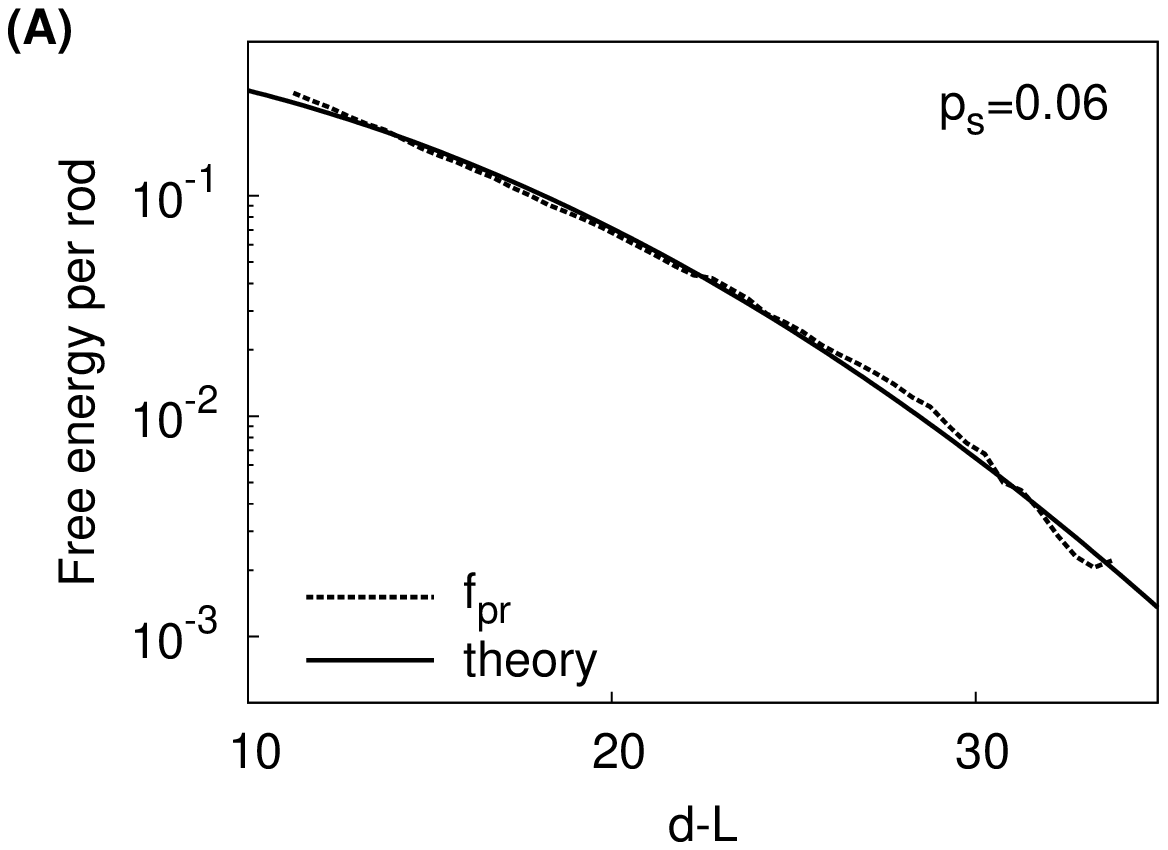}
\epsfxsize=0.49\columnwidth\epsfbox{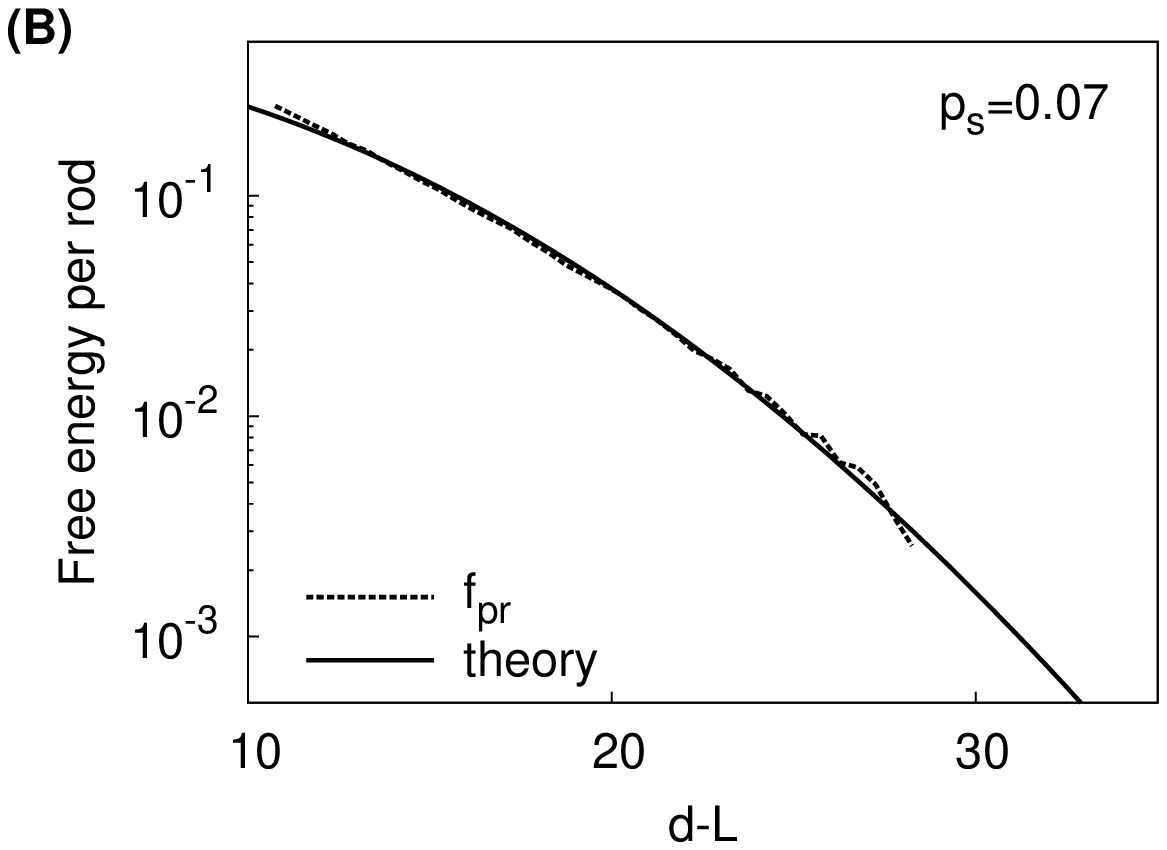}
\caption{The protrusion interaction potential is well-fit by the theory in some parameter ranges. The dotted lines show the repulsive interaction potential $\fp$ measured from simulations and the solid lines correspond to the best fit to the protrusion undulation potential given in the text with $B$ and $\gamma$ as fit parameters. Parameters are $L=100$, $\delta=1.5$ and (A) $\posm=0.06$, (B) $\posm=0.08$ and the best fit values are (A) $B=0.8$, $\gamma^{-1}=213$, (B) $B=0.9$, $\gamma^{-1}=156$.
 \label{fig:protFitGaussian}
}
\end{figure}

\emph{Origins of monolayer stability.}
To understand the nature of the repulsive membrane-membrane interactions, we determine their functional form $f_\text{pr}$ by subtracting the depletion interaction $f_\text{d}$ from the measured membrane-membrane free energy,  $f_\text{pr}(\ds)=f(d)-f_\text{d}(d)$.  The depletion term is given by $f_\text{d}(d)=\posm\langle v_\text{ex}\rangle_{d_{\text{s}}}$, where $\vex$ is the volume excluded to spheres by rods, and $\langle \cdot \rangle_{d}$ indicates an ensemble average over configurations at a particular separation $d$. We then adapt a calculation in Ref.~\cite{Safran1994} to obtain the membrane-membrane interaction due to collective protrusions as $2\rhotwod f_\text{pr} = B \exp[-\pi \gamma(d-L)^2/ 3\kt]$ with $\gamma$ the surface tension, $\rhotwod$ the area per rod, and $B$ a constant. The calculations are presented in further detail in the SI.  As shown in Fig.~\ref{fig:protFitGaussian}, the measured repulsive interaction $f_\text{pr}$ is well described by this functional form, with  fit values of $\gamma$ that are close to the surface tension extracted from simulated height-height correlation spectra (SI Fig. 1). Thus, the membrane-membrane repulsion primarily arises from collective protrusion undulations.

\emph{Experimental phase diagram.} Simulations predict a critical depletant size above which isolated membranes are unstable with respect to stacks of membranes for all osmotic pressures. We experimentally verify this prediction using a mixture of {\it fd} virus and non-adsorbing polymers(for methods see the SI). As shown in \ref{fig:phasesSphereDiameter}, there is qualitative agreement between simulations and experiments in two respects. First,  colloidal membranes are unstable for depleting polymer of large size; i.e. there is a direct transition from the nematic phase to smectic filaments. In contrast, for smaller polymer sizes, colloidal membranes are stable. Second, with decreasing polymer size the osmotic pressure (polymer concentration) at the transition from colloidal membranes to smectic filaments increases. Several points need to be considered when comparing the experimental and computational phase diagrams. First, there is a gap in the data between the polymer sizes corresponding to $R_g=9.7nm$ and $R_g=17.9nm$ due to limited commercial availability of polymers with appropriate size. Second, the transition pressure from the nematic/isotropic phase to colloidal membranes increases precipitously for smaller polymer sizes ($\Rg \lesssim 5.2$ nm). This is due to the deviations of the {\it fd} system from an ideal model hard rod system due to its surface charge. Making the depleting polymer size smaller than the electrostatic repulsion length greatly reduces the strength of the attractive interactions, requiring a higher depletant concentration to induce condensation of colloidal membranes~\cite{Dogic2004}. Third, while the chirality of the individual viruses can influence the assembly pathways, we have determined that the locations of transitions in the experimental phase diagram are independent of the chirality of the constituent rods.

In summary, this study demonstrates for the first time that entropic forces are sufficient to stabilize monolayer colloidal membranes at equilibrium. We find that collective protrusion undulations are the primary force that  stabilizes isolated membranes. While experimental observations of protrusions in Ref.~\cite{Barry2010} were inferred as individual rods, those experiments only fluorescently labeled a small fraction of rods and thus could not resolve collective modes.

The simulations also predict that the width of the isolated membrane phase depends strongly on aspect ratio and depletant size. While most previous simulations of hard rods considered small aspect ratios, our prediction of a critical aspect ratio below which the colloidal membrane phase disappears suggests that large aspect ratios are crucial for the phase behavior observed in  Ref.~\cite{Barry2010}. The predicted critical aspect ratio is only qualitative, but can be tested by monitoring the phase behavior of depletant and rods with varying lengths, as the prediction of a critical depletant size was tested here.

{\bf acknowledgement} This work was supported by NSF-MRSEC-0820492, NSF-DMR-0955776, NIH-R01AI080791, and ACS-PRF 50558-DNI7. We thank Robijn Bruinsma for insightful discussions about collective protrusion interactions.

\bibliographystyle{apsrev4-1}

\end{document}